# Origin and diversification of a metabolic cycle in oligomer world

Tomoaki Nishio, Osamu Narikiyo*

*Department of Physics, Kyushu University, Fukuoka 812-8581, Japan*

ABSTRACT

Based on the oligomer-world hypothesis we propose an abstract model where the molecular recognition among oligomers is described in the shape space. The origin of life in the oligomer world is regarded as the establishment of a metabolic cycle in a primitive cell. The cycle is sustained by the molecular recognition. If an original cell acquires the ability of the replication of oligomers, the relationship among oligomers changes due to the poor fidelity of the replication. This change leads to the diversification of metabolic cycles. The selection among diverse cycles is the basis of the evolution. The evolvability is one of the essential characters of life. We demonstrate the origin and diversification of the metabolic cycle by the computer simulation of our model. Such a simulation is expected to be the simplified demonstration of what actually occurred in the primordial soup. Our model describes an analog era preceding the digital era based on the genetic code.

* Corresponding author.
   *E-mail address:* narikiyo@phys.kyushu-u.ac.jp (O. Narikiyo)

# 1. Introduction

Various theoretical models have been proposed for the origin of life (Popa, 2004). Such models are roughly classified into two categories (Shapiro, 2007). One is the replicator-first model and the other is the metabolism-first model. Anyway, both activities should be integrated in a compartment. Thus three important conditions, replication, metabolism and compartment, should be cleared for the origin of life (Szathmary, 2006).

The metabolism condition is the first priority in the protein-world hypothesis and the compartment condition in the lipid-world hypothesis (Luisi, 2006). The RNA-world hypothesis is proposed to clear both replication and metabolism conditions at the same time (Atkins et al., 2011). However, one single hypothesis is not enough to clear all of the three conditions.

In this paper we propose a version of the oligomer-world hypothesis where the interplay between oligo-peptides and oligo-nucleotides sustains the metabolism and the replication. The original version of this hypothesis was introduced by Shimizu (Shimizu, 1996) and here we extend it. While in the RNA world both replication and metabolism are sustained solely by RNAs, in the oligomer world both proto-RNAs and proto-proteins are necessary to sustain the activities. It should be noted that some appropriate compartment is assumed but not implemented explicitly in our model as in the case of the RNA-world model.

Before discussing above hypotheses we have to clear a precondition that oligomers are supplied under a pre-biotic environment. Although it should be discussed as an issue in the study of a chemical evolution, we assume the existence of oligomers as the starting condition of our study. Such an assumption is supported by recent studies (Atkins et al., 2011). Thus our issue is to establish a system using prepared elements, oligomers. The preparation is the event of necessity but the establishment is the one of chance as discussed in the following.

The molecular recognition between oligo-peptides and oligo-nucleotides is the key ingredient of our model. Such an interplay is effectively represented in a shape space (Perelson and Weisbuch, 1997) and we also utilized the shape space to represent the molecular recognition between antigens and antibodies in the study of immune network (Saito and Narikiyo, 2011). This type of recognition is an analog process compared with a digital process like the translation of the genetic code. The translation is performed under the molecular recognition in a digital manner based on the Watson-Crick pairing. However, such a digital process needs complex molecular machines like ribosome and tRNA. Thus we assume that the era of analog molecular recognition preceded the era of digital molecular recognition. Our study only employs the analog molecular recognition.

In a shape space we construct an abstract metabolic cycle whose reactants and products are oligo-peptides and which is regulated by oligo-nucleotides. Since protein and RNA can mimic each other as seen in vitro, e.g. SELEX experiments (Atkins et al., 2011), and in vivo, e.g. the termination of the translation (Ito et al., 2000), such a distinction is introduced for convenience. In a version of the hyper-cycle model a similar relation, where the self-replication cycles of RNAs are linked by enzymatic activities of proteins to form a hyper-cycle, is assumed (Eigen, 1992).

In developmental biology the development of cells is described in the epigenetic landscape (Waddington, 1957) where stable states arise regulated by genes. The change in the regulation leads to an evolution of cells. Both in development and in evolution the regulation by genes is the most important fact. Our metabolic cycle might be the most primitive version of such regulated states.

First we demonstrate how a stable metabolic cycle that is necessary to establish a self of a cellular life arises. Next we demonstrate how diverse cycles are derived from an original cycle. Such a diversification is necessary to introduce the evolution. The evolvability is one of the essential characters of life. The present-day well-organized cells are expected to descend from the evolvable primitive cells.

## 2. Origin of metabolic cycle

We consider the origin of a metabolic cycle in the shape space (Perelson and Weisbuch, 1997). As a starting condition we assume that oligo-peptides and oligo-nucleotides are supplied as the consequence of a chemical evolution (Atkins et al., 2011). While these two kinds of oligomers are explicitly implemented in our model, the other materials are assumed to give some suitable environment to these. For example, the lipids forming a compartment, the proto-ATPs or minerals supplying energy, and the organic molecules, other than these oligomers, taking parts in metabolic reactions are implicitly included in the environment. Although both oligomers can play the same roles, we assume for convenience that the reactants and products of metabolic reactions are oligo-peptides and that such reactions are regulated by enzymatic activities of oligo-nucleotides. Here we only take such regulated reactions into account and can neglect the reactions without enzymes. Without such regulations a reaction network crashes down to tar by spreading side-reactions (Szathmary, 2006).

The coordinate of the shape space (Perelson and Weisbuch, 1997) represents some physico-chemical property of oligo-peptides. While a position in the shape space has a strong correlation to the shape of the oligo-peptide, it also reflects the information of the charge distribution, the hydrophobicity, and so on. Hereafter we employ 2-dimensional shape space for simplicity of the implementation, though the complete shape space is a higher-dimensional one. Thus an oligo-peptide is characterized by a circle in our 2-dimensional shape space. The center of the circle roughly corresponds to the frozen shape of the oligo-peptide and the radius represents the deformation under the fluctuating environment. On the other hand, an oligo-nucleotide is characterized by a rectangle in the same shape space. The rectangle is defined as the oligo-nucleotide can bind with the oligo-peptide whose characterizing circle has an overlap with the rectangle. The molecular recognition represented in this manner is an analog process.

We prepare the 2-dimensional shape space as shown in Fig. 1 where the center of an oligo-peptide is represented by the coordinate $(x, y)$ and the range is defined as $0 < x < 10$ and $0 < y < 10$.

Here we simulate the origin of a metabolic cycle by chance. Such a cycle is the minimum metabolic cycle in the oligomer world.

First we distribute $P$ oligo-peptide centers randomly in the shape space. The default value of $P$ is set as $P = 10$. The default radius of the circle $r$ is set as $r = 1$. Next we distribute $N$ oligo-nucleotide centers randomly. The default value of $N$ is set as $N = 40$. The length $l$ of a side of a rectangle is given by a uniform random number in the range of $s < l < s + 1$. The default value of $s$ is set as $s = 1$. Thus we obtain a set of oligomers assumed to be enclosed in a compartment by chance. We call such a set as a sample. For a fixed set of parameters 10000 samples are made randomly.

A sample is assumed to be enclosed in a membrane so that we call it a cell. Experimentally such a sample is easily prepared, e.g. , by the inverted emulsion method (Pautot et al., 2003) and we expect that the similar process has occurred in the primordial soup.

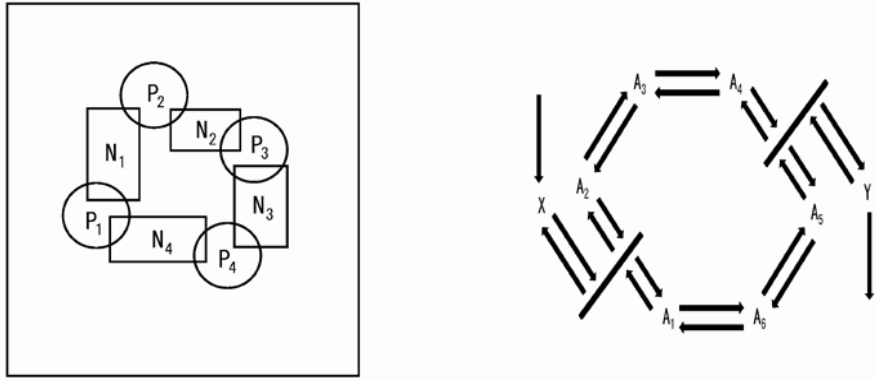

**Fig. 1.** (Left) An example of the metabolic cycle represented in 2D shape space. The overlap between $P_i$ and $N_j$ represents the binding between the oligo-peptide and the oligo-nucleotide via molecular recognition. (Right) An example of the metabolic cycle represented by chemical reactions. Such a cycle is adopted as the core of the chemoton model (Ganti, 2003). $A_k$ represents one of the reactants or the products of the chemical reaction. The raw material X is imported from the outside of the cell and the waste product Y is exported to the outside. At the junctions drawn by the diagonal lines X and Y couple to the metabolic cycle.

Most of the samples, cells, cannot form a metabolic cycle defined below but some small number of samples can by chance. The criteria for the formation of a metabolic cycle are (i) the linkage condition and (ii) the in-out condition.

The linkage condition is satisfied if such a sequence as $P_1$-$N_1$-$P_2$-$N_2$-$P_3$-$N_3$-$P_4$-$N_4$-$P_1$ is found where $P_i$ represents an oligo-peptide and $N_j$ an oligo-nucleotide, and the existence of the overlap between $P_i$ and $N_j$ is represented by $P_i$-$N_j$. The link $P_i$-$N_j$-$P_k$ represents that the reaction transforming the reactant $P_i$ into the product $P_k$ is regulated by the enzyme $N_j$. Actually our model cannot specify the direction of the reaction so that the roles of reactant and product are arbitrary. To form a cycle the first oligo-peptide in the sequence should coincide with the last oligo-peptide. An example of the cycle is shown in Fig. 1.

The in-out condition is satisfied if two points (3,3) and (7,7) are covered by some oligo-peptides belonging to the cycle defined by the sequence satisfying the linkage condition. The oligo-peptide covers the point (3,3) is assumed to take necessary material in the metabolic cycle. The oligo-peptide covers the point (7,7) is assumed to take unnecessary material out of the metabolic cycle. These points correspond to the junctions related to the necessary material (X) and the unnecessary material (Y) in the chemoton model (Ganti, 2003). This condition ensures that the cell, as an open system, can utilize the energy flow through these materials. See Fig. 1 for the core metabolic cycle in the chemoton model.

Out of randomly synthesized 10000 samples we obtain 10 cells which have established metabolic cycles satisfying the linkage and in-out conditions. Among them 7 cells have about 30 kinds of cycles. One of them has 6 kinds of cycles as shown in Fig. 2. The other two have more cycles, 250 and 294, and one of them is shown in Fig. 2.

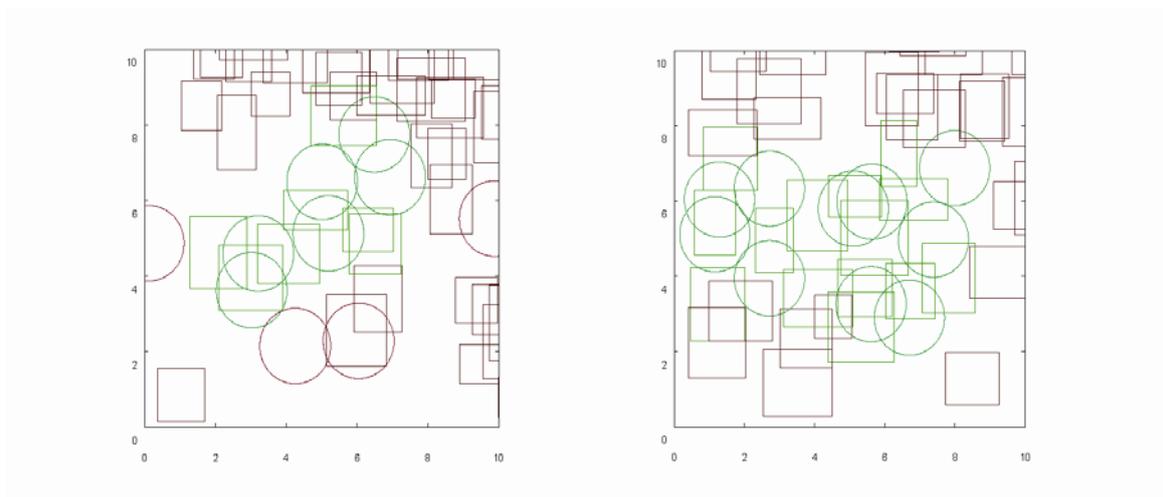

**Fig. 2.** Examples of cells with established metabolic cycles in the shape space. The objects (circles and rectangles) in green form the cycles but the objects in red do not. (Left) A simple cell where 6 oligo-peptides and 7 oligo-nucleotides form 6 kinds of cycles. (Right) A complex cell where 250 kinds of cycles are formed.

The number of cells with established metabolic cycles, out of 10000 samples, increases when we increase the number of oligo-peptides $P$ or oligo-nucleotides $N$ as shown in Fig. 3. It also increases when we increase the radius of oligo-peptides $r$ or the size of oligo-nucleotides $s$ in the shape space as shown in Fig. 4.

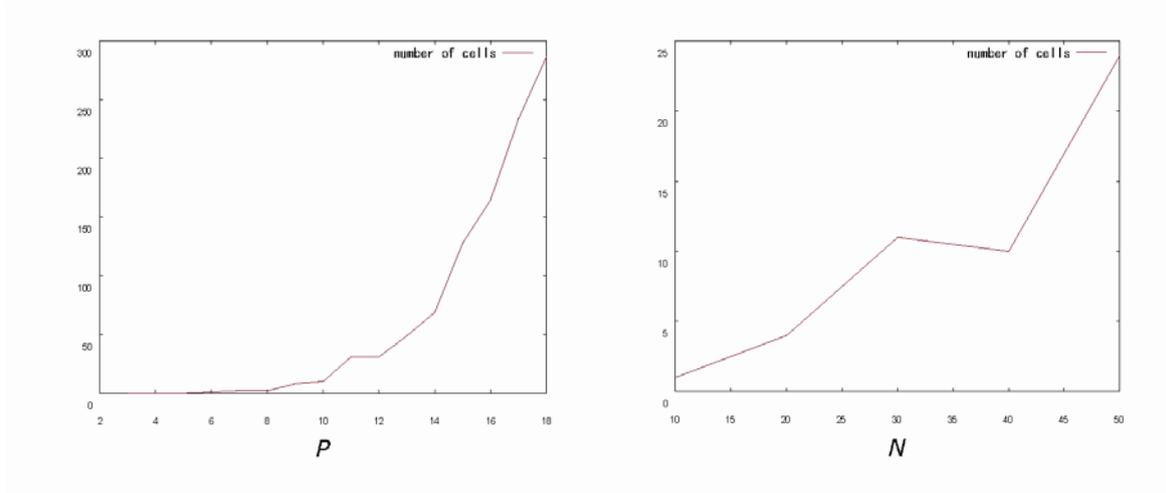

**Fig. 3.**  The number of cells with established metabolic cycles out of 10000 cells as a function of the number of oligo-peptides $P$ in a cell or the number of oligo-nucleotides $N$ in a cell. $N=40$ in the left simulation and $P=10$ in the right simulation.

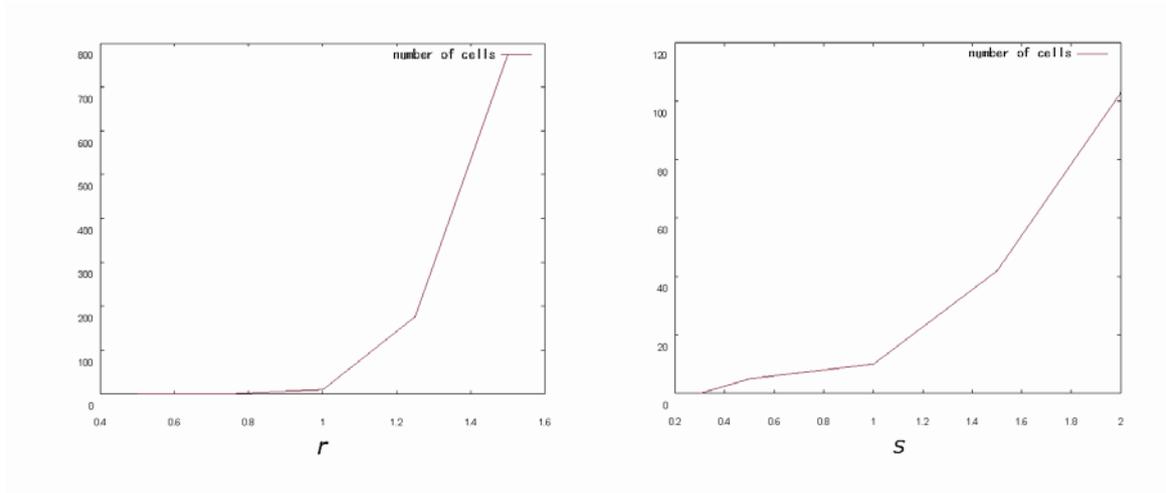

**Fig. 4.**  The number of cells with established metabolic cycles out of 10000 cells as a function of the radius of oligo-peptides $r$ or the size of oligo-nucleotides $s$ in the shape space. The parameters other than $r$ or $s$ in the horizontal axis are kept as the default values.

Since our shape space is a low-dimensional one, we can observe the origin of a metabolic cycle. Although actual shape space for oligomers must be higher-dimensional one, whole shape of an oligomer is not relevant but only some local part is relevant for the docking with the other oligomers so that the relevant shape space is expected to be relatively low-dimensional one. Thus the probability for establishing a cycle in a cell is not vanishingly small.

The cycle is formed by oligomers enclosed in a compartment by chance.

## 3. Diversification of metabolic cycle

In the preceding section we have demonstrated the origin of the minimum metabolic cycle in the oligomer world. Next in this section we extend the model and demonstrate how diverse cycles are derived from an original cycle. Such a diversification is necessary for cells to be evolvable.

In order to introduce the diversification we incorporate the replication process of oligo-nucleotides into the model. By the replication we can discuss the genetic behavior of cells where a mother cell is assumed to be divided into daughter cells.

As in a version of the hyper-cycle model (Eigen, 1992) we assume that the replication of oligo-nucleotides are supported by oligo-peptides. Thus we prepare a cell with four kinds of necessary oligo-peptides as the initial condition. Two of these to satisfy the in-out condition are the same as those in the model in the previous section. The other two support the replication of oligo-nucleotides.

The replication in our model is assumed to proceed as follows. First monomers or oligomers of nucleotides supplied from the environment form the Watson-Crick pairs on the template oligo-nucleotide in the cell. Next the ligation among nucleotides on the template occurs supported by an oligo-peptide. Hereafter we call the oligo-peptide the ligase. By the enzymatic activity of the ligase a copy of the template is obtained but it still binds to the template. Finally the copy should be removed from the template. The removal is achieved by an oligo-peptide. Hereafter we call the oligo-peptide the helicase. Then an isolated copy, which can regulate the metabolic cycle and can be the template of the replication, is obtained. Both the ligase and the helicase are necessary for the present living cells. Although the present ligase and helicase are complex, here we have assumed the presence of primitive ones with poor ability. Since the interaction forming the Watson-Crick pairs is much stronger than those for the molecular recognitions of ligase and helicase, the timescale for the formation of pairs is much shorter than those for the molecular recognitions. While the ligase can act as an enzyme without the external energy-supply, the helicase can act as a molecular motor with the supply such as proto-ATP. Thus the timescale for the ligase is much shorter than that for the helicase. Since the three timescales are separate, the replication is assumed to proceed in the above-mentioned order. We do not implement the process of the replication but assume that the existence of the ligase and the helicase in the cell ensures the replication.

Since we are interested in a primitive cell so that the mechanism of the translation from proto-RNAs (oligo-nucleotides) to proto-proteins (oligo-peptides) is absent, oligo-peptides are assumed to be supplied from the environment and not evolvable. The translation is performed under the molecular recognition in a digital manner, based on the Watson-Crick pairing, with complex molecular machines like ribosome and tRNA. On the other hand, our study only employs the molecular recognition in an analog manner as discussed in the previous section.

As an artificial initial condition we prepare the cell shown in Fig. 5 (Left). The four circles centered at (3,3), (7,7), (3,7) and (7,3) with the radius $r=1$ represents necessary oligo-peptides. The oligo-peptides centered at (3,3) or (7,7) play the same role as in the previous section and the existence of these ensures the in-out condition. The oligo-peptide centered at (3,7) acts as the ligase. The oligo-peptide centered at (7,3) acts as the helicase. The four circles centered at (0.5,5), (5,0.5), (5,9.5) and (9.5,5) with the radius $r=1$ represents the other oligo-peptides which are unnecessary in the beginning but might be involved into the modified metabolic cycle. The modification is led by the change in the oligo-nucleotides which regulate the metabolic cycle. The change is the consequence of the poor fidelity of the replication in primitive cells. Thus the diversification of metabolic cycles is introduced by the error in the replication.

Each oligo-peptide is prepared in eightfold so that there are 64 oligo-peptides in the initial cell. On the other hand, four kinds of oligo-nucleotides are prepared in fourfold so that there are 16 oligo-nucleotides. These oligo-nucleotides are represented by the rectangles shown in Fig. 5 (Left) which are centered at (3,5), (5,3), (5,7) and (7,5) and whose side lengths are 1 and 2.

The time-evolution of the initial cell is carried out as follows. If all the four necessary oligo-peptides are involved in some metabolic cycle, then the number of each oligo-peptide involved in the cycle, including unnecessary oligo-peptides, is increased by $n$ and the copy of each oligo-nucleotide is created probabilistically. These procedures define the unit time-step of our simulation. The default value of $n$ is set as $n=1$.

The criteria for the formation of a metabolic cycle are (i) the linkage condition, (ii) the in-out condition and (iii) the ligase-helicase condition. The conditions (i) and (ii) are the same as those in the previous section. The condition (iii) is parallel to (ii). Namely, the ligase-helicase condition is satisfied if two points (3,7) and (7,3) are covered by some oligo-peptides belonging to the cycle defined by the sequence satisfying the linkage condition.

The probabilistic creation of the copies is implemented as follows. Since we assume that the replication is much slower than the metabolic reactions and the formation of the Watson-Crick pairs is under the influence of the fluctuating environment, the probability $p$ to succeed in making a copy is set to be small, $p=0.08$ as the default value, and the fidelity of the copy is controlled by random numbers. Namely, a corner point of the rectangle representing the copy moves from that of the template by the random displacement $(x, y)$ where $x$ and $y$ are uniform random numbers in the ranges of $-0.5 < x < 0.5$ and $-0.5 < y < 0.5$. At the same time the other corner point in diagonal direction of the rectangle moves by another random displacement. If the resultant copy is highly deformed, $l < 0.1$ or $l > 6$ where $l$ is the length of a side of the rectangle originally set to be 1 or 2 in the initial condition, is expected to become unstable so that we delete it.

At the end of each time-step defined above we check the total number of oligo-peptides and oligo-nucleotides in the cell. If the number exceeds the critical value $N$, we put every oligomers probabilistically into two groups corresponding to daughter cells. This procedure implements the cell division in terms of the simulation rule. After the division we pursue only one of the daughter cells and repeat the above procedures for a single cell. For each oligomers the probability to be included in the daughter cell pursued is 1/2. The critical value for the cell division is chosen as $N = 300$.

The simulation starts with the artificial initial condition shown in Fig. 5 (Left) where 4 kinds of oligo-peptides and 4 kinds of oligo-nucleotides form a single cycle.

Due to the poor fidelity of copying oligo-nucleotides, the positions of the copies deviate from the initial ones as shown in Fid. 5 (Right) where only initial 4 kinds of oligo-peptides take parts in the cycle.

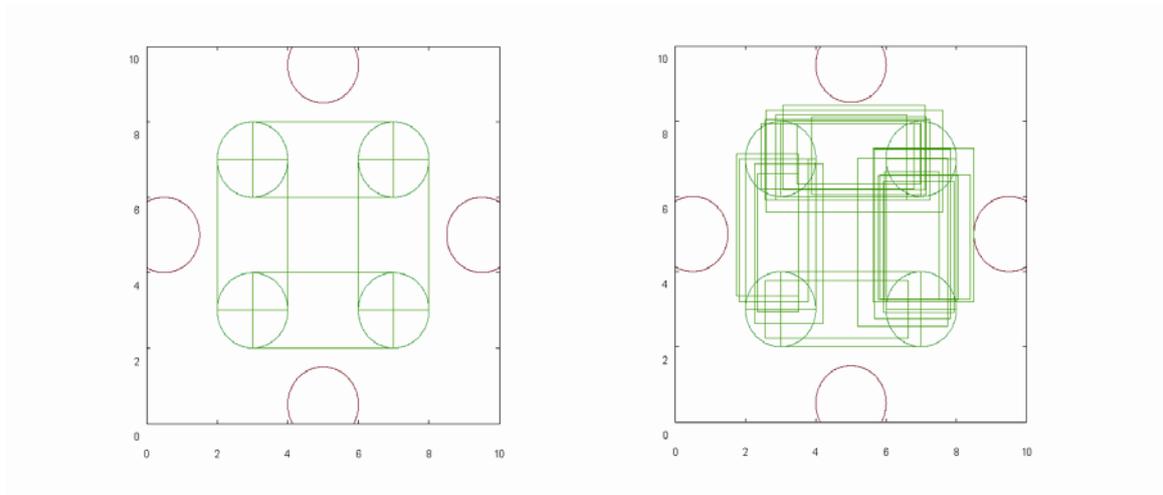

**Fig. 5.** Snapshots of a cell in the shape space. The objects (circles and rectangles) in green form the cycles but the objects in red do not. (Left) Snapshot at 0 step. (Right) Snapshot at 10 steps.

In Fig. 6 (Left) an oligo-peptide begins to take part in the cycles. It was not recognized by any oligo-nucleotides in the initial condition but is bridged to the other oligo-peptides by deformed oligo-nucleotide.

In Fig. 6 (Right) another oligo-peptide begins to take part in the cycles so that 6 kinds of oligo-peptides are involved in the cycles. On the other hand, an oligo-nucleotide, which does not bridge any pairs of oligo-peptides, appeared.

At the end of 20 time-steps there are 149 oligo-peptides and 69 oligo-nucleotides in the cell and the division of the cell has not occurred yet.

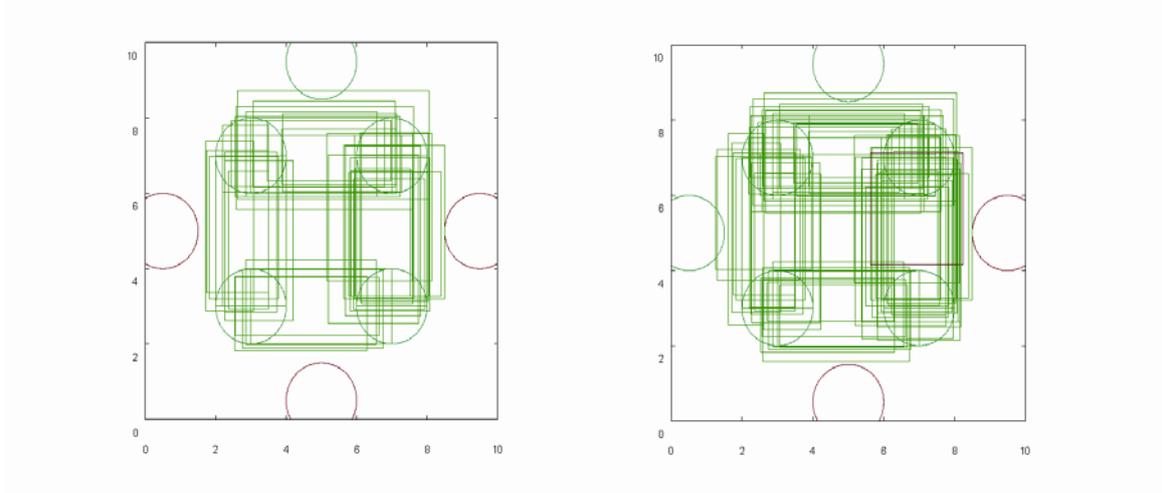

**Fig. 6.**  (Left) Snapshot at 15 steps. (Right) Snapshot at 20 steps.

In Fig. 7 (Left) the kinds of oligo-peptides involved in the cycles has increased up to 7. On the other hand, the number of oligo-nucleotides, which does not bridge any pairs of oligo-peptides, increases.

At the end of 26 time-steps the first cell division occurs. One of the daughter cells is shown in Fig. 7 (Right) where one kind of oligo-peptide has been dropped off from the cycles compared with Fig. 7 (Left). Such a drop-off occurs since the bridging oligo-nucleotides between it and the other oligo-peptides have been lost by the cell division.

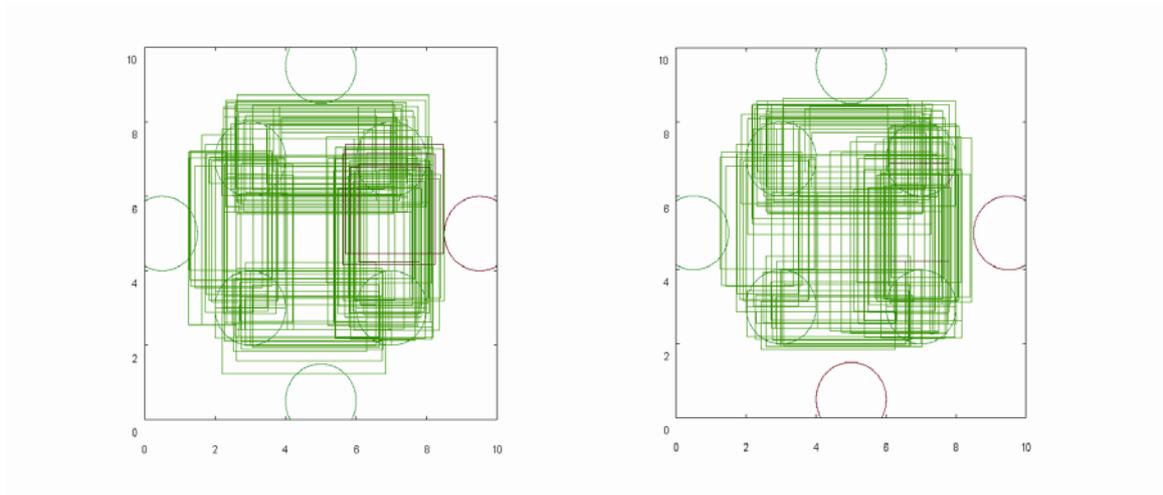

**Fig. 7.**  (Left) Snapshot at 25 steps. (Right) Snapshot at 30 steps.

At the end of 38 time-steps the second cell division occurs. One of the daughter cells of this division is shown in Fig. 8 (Left) where 7 kinds of oligo-peptides are involved in the cycles as the original mother cell before the first cell division shown in Fig. 7 (Left). Thus the descendant cells have acquired some robustness against the perturbation of the cell division. The distribution of oligo-nucleotides has become wider.

At the end of 100 time-steps the number of oligo-nucleotides, which does not bridge any pairs of oligo-peptides, has become appreciable. At the same time the kinds of oligo-peptides involved in the cycles has decreased down to 5.

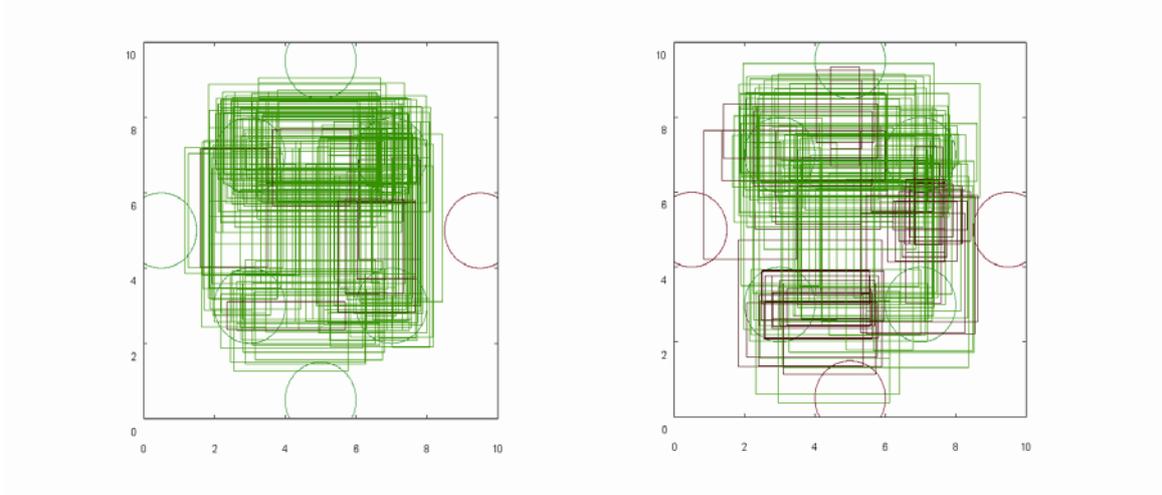

**Fig. 8.**   (Left) Snapshot at 45 steps. (Right) Snapshot at 100 steps.

The above sequence continues up to 915 time-steps until which there is at least one kind of cycle in the descendant cells. The last descendant cell with the cycle experienced 94 times cell division.

The period of the survival of the descendant cells changes when the rapidity of oligomer production $p$ or $n$ is changed as shown in Fig. 9. These data tell that the proper ratio of the oligomer numbers (# of oligo-peptides / # of oligo-nucleotides) is needed for the robustness against the cell division. If the ratio becomes large out of the proper range, the shortage of oligo-nucleotides makes the formation of the cycles difficult. If the ratio becomes small, the shortage of oligo-peptides causes the difficulty.

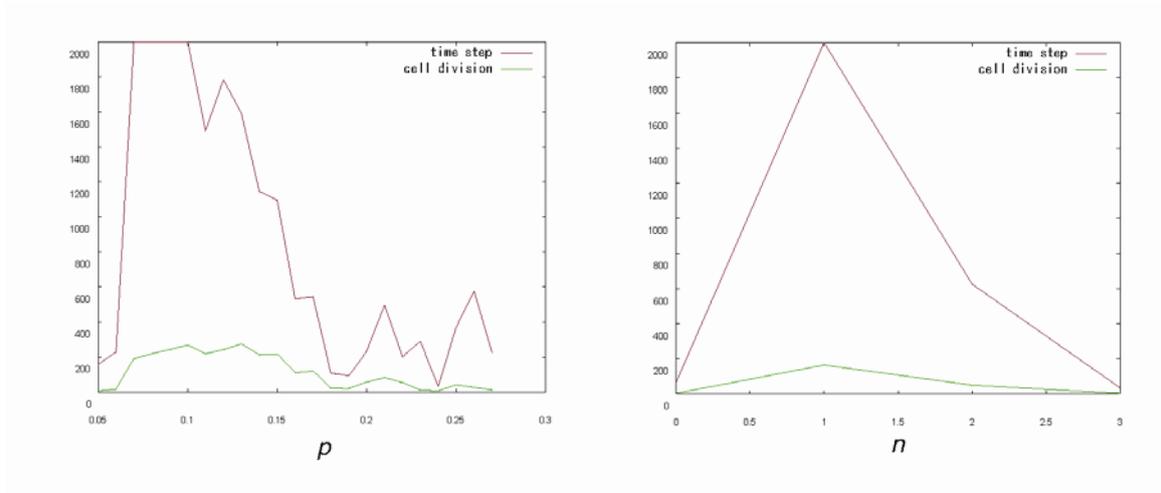

**Fig. 9.** The line graph in red represents the number of time step experienced by the cell before metabolic cycles become absent as a function of $p$ or $n$. The line graph in green represents the number of cell divisions experienced in the same period. Here $p$ is the probability of making copies of oligo-nucleotides and $n$ is the number of increase in oligo-peptides per time step. The parameters other than $p$ or $n$ in the horizontal axis are kept as the default values.

We have traced a sequence of the descendant cells with the replication process of oligo-nucleotides. The metabolic cycles in the descendant cells become diverse due to the poor fidelity of the replication.

## 4. Conclusion

We have proposed an abstract model in the shape space where the molecular recognition among oligomers sustains the metabolic cycles in the primitive cell.

The establishment of the metabolic cycle can be identified as the origin of life in the oligomer world. Oligo-peptides, oligo-nucleotides, lipids, proto-ATPs and so on are expected to be prepared through the chemical evolution under the pre-biotic environment. The preparation is the event of necessity. In the prepared primordial soup, vesicles are formed randomly and some of them can have metabolic cycles by chance. We have demonstrated the establishment in the shape space. The origin of life in our model is the event of chance.

Next we have demonstrated the diversification of metabolic cycles as the consequence of the poor fidelity of the replication. The selection among diverse cycles is the basis of the evolution. The evolvability is one of the essential characters of life. The present-day well-organized cells are expected to descend from the evolvable primitive cells.

Our model describes an analog era preceding the digital era based on the genetic code and our simulation is expected to be the simplified demonstration of what actually occurred in the primordial soup.